
\input harvmac

\def\bar{\overline}

\def\ln{{\rm ln}}
\def\exp{{\rm exp}}

\def\a{\alpha} \def\b{\beta}  
\def\f{\phi}
\def\t{\theta}  \def\p{^\prime} \def\l{\lambda}
\def\T{\Theta}

\def\ts{\thinspace}

\def\strutdepth{\dp\strutbox} 
\def\probsymbol{\vtop to \strutdepth{\baselineskip\strutdepth
  \vss\llap{\bf ??~~}\null}}
\def\prob#1{\strut\vadjust{\kern-\strutdepth\probsymbol}{\bf #1}}
\def\B{\bullet} \def\W{\circ}
\def\({\left(} \def\){\right)}
\def\[{\left[} \def\]{\right]}

%
%
\def\RF#1#2{\if*#1\ref#1{#2.}\else#1\fi}
\def\NRF#1#2{\if*#1\nref#1{#2.}\fi}
\def\refdef#1#2#3{\def#1{*}\def#2{#3}}
%
%
\def \ts{\thinspace}
\def \AJM{{\it Am.\ts J.\ts Math.\ts}}
\def \CMP{{\it Comm.\ts Math.\ts Phys.\ts }}
\def \NP{{\it Nucl.\ts Phys.\ts }}
\def \PL{{\it Phys.\ts Lett.\ts }}
\def \PR{{\it Phys.\ts Rev.\ts }}
\def \Tahoe{Proceedings of the XVIII
 International Conference on Differential Geometric Methods in Theoretical
 Physics: Physics and Geometry, Lake Tahoe, USA 2-8 July 1989}
\def \Tahoe{Proceedings of the NATO
 Conference on Differential Geometric Methods in Theoretical
 Physics, Lake Tahoe, USA 2-8 July 1989 (Plenum 1990)}
\def \Zm{Zamolodchikov}
\def \AZm{A.\ts B.\ts \Zm}
\def \AlZm{Al.\ts B.\ts \Zm}
\def\ped{P.\ts E.\ts Dorey}
\def\dur{H.\ts W.\ts Braden, E.\ts Corrigan, \ped\ and R.\ts Sasaki}
%
%
\refdef\rAa\Aa{V.\ts I.\ts Arnold, {\it Catastrophe Theory} (Springer-Verlag
 1984)}

\refdef\rAFZa\AFZa{A.\ts E.\ts Arinshtein, V.\ts A.\ts Fateev and
 \AZm, \PL {\bf B87} (1979) 389}

\refdef\rAGGa\AGGa{D.\ts Amit, Y.\ts Goldschmidt and G.\ts Grinstein,
 {\it J. Phys.} {\bf A13} (1980) 585}

\refdef\rBa\Ba{H.W. Braden, \lq A note on affine Toda couplings', Edinburgh
preprint 91-01}

\refdef\rBb\Bb{See for example R. Baxter,
 {\it Exactly Solved Models in Statistical Mechanics} (Academic Press 1982)}

\refdef\rBc\Bc{N.\ts Bourbaki, {\it Groupes et alg\`ebres de Lie} {\bf
 IV, V, VI,} (Hermann, Paris 1968)}

\refdef\rBBSSa\BBSSa{F. A. Bais, P. Bouwknegt, M. Surridge and K. Schoutens,
 \NP {\bf B304} (1988) 348, 371}

\refdef\rBCa\BCa{R. Brunskill and A. Clifton-Taylor, {\it English Brickwork}
 (Hyperion 1977)}

\refdef\rBCDSa\BCDSa{\dur, \PL {\bf B227} (1989) 411}

\refdef\rBCDSb\BCDSb{\dur, \Tahoe}

\refdef\rBCDSc\BCDSc{\dur, \NP {\bf B338} (1990) 689}

\refdef\rBCDSd\BCDSd{\dur,
 ``Aspects of affine Toda field theory", to appear in the Proceedings
of the $10^{\rm th}$ Winter School on Geometry and Physics, Srni,
Czechoslovakia; Integrable Systems and Quantum Groups, Pavia, Italy;
Spring Workshop on Quantum Groups, ANL, USA}

\refdef\rBCDSe\BCDSe{\dur, \NP {\bf B356} (1991) 469}

\refdef\rBCGTa\BCGTa{E. Braaten, T. Curtright, G. Ghandour and C. Thorn, {\it
 Phys. Lett}. {\bf B125} (1983) 301}

\refdef\rBGa\BGa{P. Bowcock and P. Goddard, \NP {\bf B285} (1987) 651}

\refdef\rBLa\BLa{D. Bernard and A. LeClair, \lq\lq Quantum group
 symmetries and non-local currents in 2D QFT", preprint CLNS-90/1027,
 SPhT-90/144}

\refdef\rBPZa\BPZa{A. A. Belavin, A. M. Polyakov and A. B. Zamolodchikov, \NP
 {\bf B241} (1984) 333}

\refdef\rBSa\BSa{H. W. Braden and R. Sasaki, \PL {\bf B255} (1991) 343}

\refdef\rBTa\BTa{D. Bernard and J. Thierry-Mieg, Comm. Math. Phys. 111 (1987)
 181}

\refdef\rCa\Ca{P. Christe, ``S-matrices of the tri-critical
 Ising model and Toda systems", \Tahoe}

\refdef\rCb\Cb{S. Coleman, {\it Aspects of Symmetry}, (Cambridge University
 Press 1985)}

\refdef\rCc\Cc{J. Cardy, in {\it Les Houches XLIX - Champs, Cordes et
 Ph\'enom\`enes Critiques} (1988)}

\refdef\rCd\Cd{S. Coleman, \PR {\bf D11} (1975) 2088}

\refdef\rCe\Ce{E. Corrigan, private communication}

\refdef\rCf\Cf{R.\ts Carter, {\it Simple Groups of Lie Type}, (Wiley
 1972)}

\refdef\rCg\Cg{H.\ts Coxeter, {\it Am. J. Math.} {\bf 62} (1940) 457}

\refdef\rCh\Ch{H.\ts Coxeter, {\it Regular Polytopes}, (Methuen 1948)}

\refdef\rCi\Ci{\Cg\semi\Ch}

\refdef\rCDa\CDa{E. Corrigan and P.E. Dorey,
\lq A representation of the exchange relation for affine
Toda field theory, Durham/Saclay preprint, September 1991}

\refdef\rCIZa\CIZa{A. Cappelli, C. Itzykson and J-B Zuber, \CMP
 {\bf 113} (1987) 1}

\refdef\rCLa\CLa{E. Corrigan and W. Lerche, private communication}

\refdef\rCMa\CMa{P.\ts Christe and G.\ts Mussardo, {\it Nucl. Phys}.
 {\bf B330} (1990) 465}

\refdef\rCMb\CMb{P.\ts Christe and G.\ts Mussardo,
 {\it Int.~J.~Mod.~Phys.}~{\bf A5} (1990) 4581}

\refdef\rCMc\CMc{J.\ts Cardy and G.\ts Mussardo, ``S-matrix of the Yang-Lee
 edge singularity in two dimensions", \PL {\bf B225} (1989) 275}

\refdef\rCMd\CMd{S. Coleman and J. Mandula, \PR {\bf 159} (1967) 1251}

\refdef\rCNa\CNa{S.\ts Coleman and R.\ts Norton, {\it Nuovo\ts Cimento} {\bf
38}
 (1965) 438}

\refdef\rCSa\CSa{E. Corrigan and R. Sasaki, private communication}

\refdef\rCTa\CTa{S.\ts Coleman and H.\ts Thun, \CMP {\bf 61}
 (1978) 31}

\refdef\rDa\Da{Vl.\ts Dotsenko, {\it Advanced Studies in Pure Mathematics} {\bf
 16} (1988) 123}

\refdef\rDb\Db{\ped, Durham PhD thesis, unpublished}

\refdef\rDc\Dc{\ped,  \NP {\bf B358} (1991) 654}

\refdef\rDd\Dd{\ped, \lq Root systems and purely elastic S-matrices II,'
preprint RIMS-787/Saclay-SPhT/91-140, August 1991}

\refdef\rDDa\DDa{C.\ts Destri and H.\ts J.\ts de Vega, {\it Phys. Lett.}
 {\bf B233} (1989) 336}

\refdef\rDSa\DSa{V.\ts G.\ts Drinfel'd and V.\ts V.\ts Sokolov,
 {\it J. Sov. Math.} {\bf 30} (1984) 1975}

\refdef\rELOPa\ELOPa{R. J. Eden, P. V. Landshoff, D. I. Olive and J. C.
 Polkinghorne,  {\it The Analytic S-matrix}, (Cambridge University Press 1966)}

\refdef\rEYa\EYa{T. Eguchi and S-K Yang, {\it Phys. Lett.} {\bf  B224} (1989)
 373}

\refdef\rFa\Fa{J. Frame, {\it Duke Math. J.} {\bf 18} (1951) 783}

\refdef\rFb\Fb{M.\ts D.\ts Freeman, \PL {\bf B261} (1991) 57}

\refdef\rFHJa\FHJa{S. Fubini, A. Hanson and R. Jackiw, \PR {\bf D7} (1973)
 1732}
\refdef\rFJa\FJa{I.B. Frenkel and N. Jing, {\it Proc. Nat. Acad. Sci.
U.S.A.} {\bf 85} (1988) 9373}

\refdef\rFKa\FKa{L. D. Faddeev and V. E. Korepin, {\it Theor. Mat. Fiz.}
 {\bf 25} (1975) 147}

\refdef\rFKb\FKb{I. B. Frenkel and V. G. Kac, {\it Inv. Math.} {\bf 62} (1980)
 23}

\refdef\rFKMa\FKMa{P. G. O. Freund, T. Klassen and E. Melzer, {\it Phys. Lett.}
 {\bf B229} (1989) 243}
\refdef\rFLOa\FLOa{A. Fring, H.C. Liao and D. Olive, \lq The mass spectrum
and coupling in affine Toda theories' IC/TP-90-91/25}

\refdef\rFLa\FLa{S. L. Lukyanov and V. A. Fateev, Lectures given in second
 Spring School \lq Contemporary Problems in Theoretical Physics', Kiev 1988,
 preprints ITF-88-74P, 88-75P and 88-76P}

\refdef\rFLMWa\FLMWa{P.\ts Fendley, W.\ts Lerche, S.\ts Mathur and N.\ts
Warner,
 \lq\lq $N=2$ supersymmetric integrable models from affine Toda field
 theories", preprint CTP-1865, CALT-68-1631, HUTP-90/A036}

\refdef\rFZa\FZa{V.\ts A.\ts Fateev and \AZm, {\it Int. J. Mod. Phys.} {\bf A5}
 (1990) 1025}

\refdef\rGa\Ga{C.\ts J.\ts Goebel, {\it Prog. Theor. Phys. Suppl.} {\bf 86}
 (1986) 261}

\refdef\rGb\Gb{P.\ts Ginsparg, in {\it Les Houches XLIX - Champs, Cordes et
 Ph\'enom\`enes Critiques} (1988)}

\refdef\rGKOa\GKOa{P. Goddard, A. Kent and D. Olive, \PL {\bf B152} (1985) 88;
 \CMP {\bf 103} (1986) 105}

\refdef\rGLPZa\GLPZa{M. Grisaru, A. Lerda, S. Penati and D. Zanon,
``Renormalization Group Flows in Generalized Toda Field Theories", MIT preprint
CTP \#1850 (1990)}

\refdef\rGNa\GNa{J-L. Gervais and A. Neveu, {\it Nucl. Phys.} {\bf B224} (1983)
 329}

\refdef\rGNOSa\GNOSa{P. Goddard, W. Nahm, D. Olive and A. Schwimmer, {\it Comm.
 Math. Phys.} {\bf 107} (1987) 179}

\refdef\rGWa\GWa{D. Gepner and E. Witten, \NP {\bf B278} (1986) 493}

\refdef\rHa\Ha{S. Helgason, {\it Differential geometry, Lie groups, and
 symmetric spaces} (Academic Press Inc. 1978)}

\refdef\rHb\Hb{T.\ts J.\ts Hollowood, \lq\lq A quantum group approach to
constructing factorizable S-matrices", Oxford preprint OUTP-90-15P}

\refdef\rHc\Hc{T.J. Hollowood, private communication}
\refdef\rHMa\HMa{T.\ts J.\ts Hollowood and P.\ts Mansfield, \PL {\bf B226}
 (1989) 73}

\refdef\rFTa\FTa{L. D. Faddeev and L. A. Takhtajan, {\it Hamiltonian methods in
 the theory of solitons}, (Springer, New York 1987)}

\refdef\rJa\Ja{M.\ts Jimbo, {\it Yang-Baxter equations in integrable systems},
 (World Scientific, 1990)}

\refdef\rJb\Jb{M.\ts Jimbo, {\it Int. J. Mod. Phys.} {\bf A4} (1989)
 3759, in {\it Braid Group, Knot Theory and Statistical Mechanics} (World
 Scientific 1989)}

\refdef\rKa\Ka{M.\ts Karowski, \NP {\bf B153} (1979) 244}

\refdef\rKb\Kb{B.\ts Kostant, \AJM {\bf 81} (1959) 973}

\refdef\rKSa\KSa{R. K\"oberle and J. A. Swieca, \PL {\bf B86}
 (1979) 209}

\refdef\rKSRa\KSRa{P. Kulish, E. Sklyanin and N. Reshetikhin, {\it
Lett. Math. Phys.} {\bf 5} (1981) 393}

\refdef\rKSRJa\KSRJa{\KSRa\semi\Jb}

\refdef\rKTa\KTa{M. Karowski and H. Thun, {\it Nucl. Phys.} {\bf B130}
 (1977) 295}

\refdef\rKMa\KMa{T.\ts R.\ts Klassen and E.\ts Melzer, {\it Nucl. Phys.}
 {\bf B338} (1990) 485}

\refdef\rLa\La{J.\ts Lepowsky, {\it Proc. Natl. Acad. Sci. USA} {\bf 82}
(1985) 8295}

\refdef\rLWa\LWa{J.\ts Lepowsky and R.L.\ts Wilson, {\it Commun. Math.
Phys.} {\bf 62} (1978) 43}

\refdef\rLSa\LSa{A. N. Leznov and M. V. Saveliev, {\it Group methods for the
 integration of nonlinear dynamical systems} (Moscow Nauka 1985)}

\refdef\rLWb\LWb{W.\ts Lerche and N.\ts P.\ts Warner, \NP {\bf B358}
(1991) 571}

\refdef\rMa\Ma{P. Mansfield, {\it Nucl. Phys.} {\bf B222} (1983) 419}

\refdef\rMb\Mb{G. Mussardo, ``Away from criticality: some results from
 the S-matrix approach", \Tahoe}

\refdef\rMCa\MCa{\Ca\semi\Mb\semi\CMb}

\refdef\rMc\Mc{S-K Ma, {\it Modern Theory of Critical Phenomena} (Benjamin
 1976)}

\refdef\rMd\Md{N.\ts J.\ts MacKay, \NP {\bf B356} (1991) 729}

\refdef\rMe\Me{N.\ts J.\ts MacKay, Rational R-matrices in irreducible
representations, to appear in J. Phys. A}

\refdef\rMf\Mf{N.\ts J.\ts MacKay, The fusion of R-matrices using the
Birman-Wenzl-Murakami algebra, Durham preprint June 1991}

\refdef\rMJa\MJa{M. Jimbo and T. Miwa, Vertex Operators in Mathematics and
Physics, eds J. Lepowsky, S. Mandelstam and I.M. Singer, MSRI {\bf\# }3
(Springer-Verlag New York 1985)}

\refdef\rMSa\MSa{G. Mussardo and G. Sotkov, preprint ``Bootstrap program and
 minimal integrable models" UCSBTH-89-64/ISAS-117-89}

\refdef\rMOPa\MOPa{A. V. Mikhailov, M. A. Olshanetsky and A. M. Perelomov, {\it
 Comm. Math. Phys.} {\bf 79} (1981) 473}

\refdef\rOa\Oa{A. Ocneanu, \Tahoe}

\refdef\rOTa\OTa{D. I. Olive and N. Turok, {\it Nucl. Phys.} {\bf B215} (1983)
 470}
\refdef\rOTb\OTb{D. I. Olive and N. Turok, {\it Nucl. Phys.} {\bf B265}
(1986) 469}
\refdef\rOTc\OTc{D. I. Olive and N. Turok, unpublished manuscript}

\refdef\rOWa\OWa{E. Ogievetsky and P. Wiegmann, {\it Phys. Lett.} {\bf B168}
 (1986) 360}

\refdef\rPa\Pa{A. M. Polyakov, {\it Phys. Lett.} {\bf B72} (1977) 224}

\refdef\rPb\Pb{G. Parisi, {\it Statistical Field Theory} (Addison-Wesley 1988)}

\refdef\rPc\Pc{V.\ts Pasquier, {\it Nucl. Phys.} {\bf B285} (1987) 162}

\refdef\rPd\Pd{A. M. Polyakov, {\it Gauge Fields and Strings} (Harwood 1987)}

\refdef\rPe\Pe{A. M. Polyakov, {\it JETP Letters} {\bf 12} (1970) 381}

\refdef\rPPa\PPa{A. Z. Patashinskii and V. I. Pokrovskii, {\it Fluctuation
 Theory of Phase Transitions} (Pergamon 1979)}

\refdef\rPTa\PTa{P. Pfeuty and G. Toulouse, {\it Introduction to the
Renormalization Group and to Critical Phenomena} (Wiley 1977)}

\refdef\rSa\Sa{G. Segal, {\it Comm. Math. Phys.} {\bf 80} (1981) 301}

\refdef\rSb\Sb{E. K. Sklyanin, \NP {\bf B326} (1989) 719}

\refdef\rSc\Sc{S. Shenker, in {\it Les Houches XXXIX - Recent Advances in Field
 Theory and Statistical Mechanics} (1982)}

\refdef\rSd\Sd{R. Sasaki, private communication}

\refdef\rSe\Se{H.\ts Samelson, {\it Notes on Lie algebras} (Van
 Nostrand 1969)}
\refdef\rSf\Sf{F.A. Smirnov, in {Introduction to Quantum Groups and
Integrable Massive Models of Quantum Field Theory}, Nankai Lectures in
Mathematical Physics, Mo-Lin Ge and Bao-Heng Zhao (eds), (World Scientific
1990)}

\refdef\rSWa\SWa{R. Shankar and E. Witten, \PR {\bf D17} (1978) 2134}

\refdef\rSZa\SZa{G. Sotkov and C-J. Zhu, {\it Phys. Lett.} {\bf B229} (1989)
 391}

\refdef\rWa\Wa{G. Wilson, {\it Ergod. Th. and Dynam. Sys.} {\bf 1} (1981) 361}

\refdef\rWb\Wb{G.\ts Watts, \PL {\bf B245} (1990) 65}

\refdef\rWc\Wc{R.S. Ward, {\it Springer Lecture Notes in Physics} {\bf 280}
(1987) 106}

\refdef\rWKa\WKa{K. Wilson and J. Kogut, {\it Phys. Rep.} {\bf 12C} (1974) 75}

\refdef\rYa\Ya{H. Yoshii, {\it Phys. Lett.} {\bf B223} (1989) 353}

\refdef\rYGa\YGa{See for example several of the articles in: C. N. Yang and M.
 L. Ge, {\it Braid Group, Knot theory and Statistical mechanics,} (World
 Scientific 1989)}

\refdef\rZa\Za{\AZm, ``Integrable Field Theory from Conformal Field Theory",
 Proceedings of the Taniguchi Symposium, Kyoto (1988)}
\refdef\rZb\Zb{\AZm, {\it Int. J. Mod. Phys.} {\bf A4} (1989) 4235}
\refdef\rZc\Zc{\AZm, {\it JETP Letters} {\bf 43} (1986) 730}
\refdef\rZd\Zd{\AZm, {\it Int. J. Mod. Phys.} {\bf A3} (1988) 743}
\refdef\rZe\Ze{\AZm, {\it Sov. Sci. Rev., Physics}, {\bf v.2} (1980)}
\refdef\rZf\Zf{\AZm, {\it Teor. Mat. Fiz.} {\bf 65} (1985) 347}
\refdef\rZh\Zh{\AZm, ``Exact Solutions of Conformal Field Theory in Two
Dimensions and Critical Phenomena", Kiev IMP preprint 87-65P (1987)}
\refdef\rZt\Zt{\AZm, Talk given in Oxford, January 1989}

\refdef\rZz\Zz{\Za\semi\Zb}

\refdef\rZg\Zg{\AlZm, ``Thermodynamic Bethe Ansatz in Relativistic Models.
 Scaling 3-state Potts and Lee-Yang Models", Moscow ITEP preprint (1989)}

\refdef\rZZa\ZZa{\AZm\  and \AlZm, {\it Ann. Phys.}
 {\bf 120} (1979) 253}


\Title{DTP/91-37---SPhT/91/108 [revised]}
{\vbox{\centerline {A representation of the exchange
relation
for}\vskip2pt\centerline{affine
Toda field theory}}}
\centerline{E. Corrigan}

\smallskip\centerline{Department of Mathematical Sciences,}
\centerline{ University of Durham, Durham DH1 3LE, UK}
\bigskip
\centerline{P.E.  Dorey}
\smallskip\centerline{Service de Physique Th\'eorique de
Saclay\rlap,\foot{{\it Laboratoire de la Direction des Sciences
de la Mati\`ere du Commissariat \`a l'Energie Atomique}}}
\centerline{91191 Gif-sur-Yvette cedex, France}
\vskip .5in

Vertex operators are constructed providing  representations of the
exchange relations containing either
the S-matrix of a real coupling (simply-laced)
affine Toda field theory, or its minimal counterpart. One feature of the
construction is that the bootstrap relations for the S-matrices follow
automatically from those for the conserved quantities, via an algebraic
interpretation of the fusing of two particles to form a single bound state.

\Date{September 1991}

\vfill\eject

\newsec{Introduction}
Stimulated by a study of perturbed conformal field
theory \NRF\rZb{\Zb\semi\FZa}\NRF\rEYa{\EYa\semi\HMa}\refs{\rZb ,\rEYa},
there has been
something of a revival of interest in two-dimensional affine Toda field
theories, whose study was begun long
ago \RF\rMOPa{\MOPa\semi\Wa\semi\OTa} but is nowhere near complete.
One of the striking features of two-dimensional integrable theories is the
possibility of making plausible guesses for their exact S-matrices on the
basis of the bootstrap and Yang-Baxter equations
\NRF\rKa\Ka\NRF\rZZa\ZZa\refs{\rKa ,\rZZa}.
For the real
coupling affine Toda field theories, based on the ADE Lie algebras,
the Yang-Baxter equation itself
plays no r\^ole because the S-matrices
are entirely diagonal. Nevertheless, the proposed
S-matrices enjoy an interesting analytic
structure as a consequence of the bootstrap alone
\NRF\rAFZa\AFZa
\NRF\rBCDSb\BCDSb\NRF\rBCDSc\BCDSc\NRF\rBCDSe\BCDSe\NRF\rBSa\BSa
\NRF\rCMa{\CMa\semi\CMb}
\NRF\rDDa\DDa\NRF\rKMa\KMa
\refs{\rAFZa {--}\rKMa}.
At least some
of this interesting structure can be seen in perturbation theory
but proofs of the conjectures are not yet available. The same bootstrap
structure appears in perturbations of certain conformal field theories, namely
the coset models
$g^{(1)}\times g^{(1)}/g^{(2)}$. However, there
the S-matrices are slightly different; while the
affine Toda S-matrices have a factor with coupling constant dependent
zeroes in the physical strip, this is
absent from the proposals for perturbed conformal field theory. In fact, it is
these \lq minimal' S-matrices, of interest in their own right, that will
be discussed first below.

In \NRF\rDc\Dc\NRF\rDd\Dd\refs{\rDc ,\rDd}, the bootstrap and
its accompanying fusing rules were found to be intimately related to the
geometry of root systems.
Moreover,  formulae for
the conjectured
S-matrices were discovered which made  clearer their structural
relationship with the roots and with the bootstrap.
The ideas might well have  more general significance given the
marked similarities between the affine Toda theory S-matrices and those
conjectured, for example,  for the
principal chiral models \RF\rOWa{\OWa\semi\Md}.
Very similar mathematical
structures have already
been observed
in the context of certain $N=2$ Landau-Ginzburg models \RF\rLWb\LWb .

The S-matrices for ADE affine Toda field theory
can be written in a number of equivalent ways. The expressions most
suitable for the present discussion will be summarised in section two,
alongside some useful facts concerning the action of the Coxeter element
of the Weyl group on the roots and weights. For more details concerning
the latter, see for example \RF\rBc{\Bc\semi\Cf\semi\Kb}.

Section three returns to an old idea in which the
S-matrix appears in a \lq braiding', or \lq exchange'
relation \NRF\rZZa\ZZa\NRF\rSf\Sf\refs{\rZZa ,\rSf}:
\eqn\braid{V_a(\t_a)V_b(\t_b)=S_{ab}(\t_a-\t_b)V_b(\t_b)V_a(\t_a),}
where each of the operators $V(\t )$ is formally associated  with a
particle of the theory,  $\t$ denoting its rapidity.

According to the present understanding of real coupling
affine Toda theory, each operator is a singlet since the particles
are distinguished by conserved charges of non-zero spin.
The assumption of associativity for the exchange
relation then has no consequence for the S-matrix---which is merely a
set of
numbers, one for each pair of particles. In
more general situations,
at least some particles will be
degenerate and the associativity of \braid\ implies the Yang-Baxter or
factorisation equation for the S-matrix.

The expressions for the S-matrices
given in section two are very suggestive and in section three
a vertex operator representation of the exchange relation will be presented,
thus  providing
a set of generating relations
for the S-matrix in a fairly natural way. The vertex
operators to be  used in this context
are reminiscent of those used by Lepowsky and Wilson
\NRF\rLWa\LWa\NRF\rLa\La\refs{\rLWa ,\rLa}\ to
obtain (twisted) representations of Kac-Moody algebras.
It appears at first sight that
the twisted vertex operators are more appropriate in the
context of affine Toda field theory than those
used by
Frenkel, Kac and Segal \RF\rFKb{\FKb\semi\Sa}
to provide the level one representations of
simply-laced affine algebras. Despite the similarities, there are
some crucial differences too,
the most important being the absence of an
obvious action of the conformal, or Virasoro, generators.
Since the field theories under discussion are massive, and
therefore not conformal, this is
hardly surprising.

Given the exchange relation \braid , it is natural to ask
about the bootstrap itself. Suppose there is an \lq operator product'
expansion of $V_a(\t_a)V_b(\t_b)$, not in the sense of a short distance
expansion but rather in the sense of a bound-state fusing relation. In
other words, suppose that, for certain (imaginary) rapidity differences,
the two-particle
state containing particles $a$ and $b$ is indistinguishable in terms of
its quantum numbers from another (on-shell)
single particle state $c$. From the point
of view of the operators, it might then be expected that there should be a
relation of the form,
\eqn\opprod
{V_a(\t_a)V_b(\t_b)\sim {c^{abc }V_{c}(\t_{c} )\over
(\t_a-\t_b-iU^c_{ab})^{n_{abc }}} ,}
where
$$\eqalign{&\t_a\sim \t_{c} -i\bar U^b_{ac}\cr
           &\t_b\sim \t_{c} +i\bar U^a_{bc},\cr}$$
and $\bar U =\pi-U$. Momentum conservation (corresponding to the first
conserved charge), together with the fact that the particles created by
the operators $V(\t )$ are always on-shell, requires that the relative
rapidity of $a$ and $b$ is just $i$ times $U^c_{ab}$, the fusing
angle for a bound state in $ab\rightarrow ab$ scattering.
It is a feature of the construction presented in section three that the
numerical factor in \opprod\ is meromorphic, the quantities $n_{abc}$
being certain integers whose properties are
described briefly at the end of that section.

Multiplying \opprod\ by $V_d(\t_d)$ and using
\braid\ leads to  the bootstrap relation between different
S-matrix elements:
\eqn\bootstrap
{S_{cd}(\t_{c}-\t_d)=S_{ad}(\t_{c}-\t_d-i\bar U^b_{ac})
S_{bd}(\t_{c}-\t_d+i\bar U^a_{bc}). }
The affine Toda field theories have infinitely many conserved charges
$P_r,P_{-r}$ where,
modulo the Coxeter number $h$, the (positive) spin-label $r$ runs over
the exponents of the
algebra defining the theory. It would be
expected that
\eqn\pcomm{[P_r,V_a(\t_a )]=p^a_re^{r\t_a}V_a(\t_a),}
where $p^a_r$ is the eigenvalue of the charge with spin $r$ on the
single type-$a$ particle state
\eqn\peigen{P_{s+kh}|p_a>=p_{s+kh}^a e^{(s+kh)\t_a}|p_a>.}
Given \pcomm , \bootstrap\ implies the
bootstrap equation for the charges, namely
\eqn\pbootstrap{p^a_re^{-ir\bar U^b_{ac}}+p^b_re^{ir\bar U^a_{bc}}=p^{c}_r.}

In this article, a formalism will be developed far enough
to provide a representation of
\braid\ and \opprod\ appropriate to the known minimal solutions of
\bootstrap\
and, after a simple modification, to real coupling
affine Toda S-matrices. Despite the absence of any derivation of these
operators from a quantisation of the original affine Toda Lagrangian, their
form is suggestive,
and they seem to provide a natural setting in
which to place the fusing rule and S-matrix formulae described in
\refs{\rDc ,\rDd}.

\newsec{Preliminaries}

The discussion will be restricted to the theories associated with the
simply-laced (ADE) series of Lie algebras.
For a given theory, each particle is
unambiguously associated to one of the
spots of the relevant Dynkin diagram. This follows from the
observation \NRF\rBCDSb\BCDSb\NRF\rFKMa\FKMa\refs{\rBCDSb ,\rFKMa}\
(now proved Lie algebraically \RF\rFb{\Fb\semi\FLOa})
that the set of classical masses of the Toda
particles can be arranged to be the components of the smallest eigenvalue
eigenvector of the corresponding (non-affine) Cartan matrix.
Besides picking a basis of simple
roots for the algebra, it also appears to be useful to divide the particles
of the field theory into two sets. This division reflects a special
property of root systems which allows the simple roots to be split
into two sets (black and white) so that within each set, all the
roots are orthogonal to one another.
The sets, the roots belonging to them, the particles themselves, and the
fundamental weights associated with the simple roots in each set
will be distinguished
wherever necessary by the symbols $\B$ or $\W$. (In
ref \rDc\ a
slightly
different notation was used, the black roots being referred to as type $\a$
the white roots as type $\b$.)

Corresponding to the
colouring of the simple roots, it is natural to choose a particular
Coxeter element of the Weyl group. Let $w_i$ be the Weyl reflection
corresponding to the simple root $\a_i$ ($i=1,\dots ,r$)  and define
$$w_\B =\prod_{i\in\B} w_i \qquad w_\W =\prod_{i\in\W} w_i$$
where the black set is
labelled $i=1,\dots ,b$ and the white set $i=b+1,\dots ,r$. Then $w$,
defined by
$$ w =w_\B w_\W ,$$
is a Coxeter element. Another subset of linearly independent roots
is defined in terms of
the set of simple roots $\a_i$ in the following way. Set
\eqn\groots{ \f_i=w_rw_{r-1}\dots w_{i+1}\a_i}
so that with the labelling introduced above
$$ \f_\W=\a_\W \quad \hbox{and}\quad \f_\B=w_\W\a_\B.$$
Moreover, if $\l_i$ denotes the fundamental weights, satisfying (for
simply-laced algebras),
$$ \l_i\cdot\a_j=\delta_{ij},$$
then the roots defined by \groots\  satisfy
\eqn\glreln{ \f_i=(1-w^{-1})\l_i.}
This formula has also been used to relate the fusing rule given in \rDc\
to the previously observed Clebsch-Gordan property of the affine Toda
couplings \RF\rBa\Ba .

It will be useful to have a (complex) basis of eigenvectors of the
Coxeter element $w$. The elements of this basis are conveniently labelled
by the exponents of the algebra. Thus, for each exponent $s$, there is an
eigenvector $e_s$ satisfying
\eqn\ebasis{w e_s=e^{2\pi is/h} e_s}
and normalised so that
\eqn\enorm{e_s\cdot e_{s^\prime}=\delta_{s+s^\prime ,h}}
where $h$ is the Coxeter number (ie order of the Coxeter element).
Occasionally, exponents $h/2$ are repeated (in the $D_{\rm even}$ series).
However, even in these cases
the same notation will suffice without confusion.

The exponents of the algebra also label the eigenvectors of the Cartan
matrix: for each exponent there is an eigenvector
\eqn\cartan{C_{ij}q_j^{(s)}=(2-2\cos\pi s/h )q_i^{(s)}}
and the eigenvectors are orthogonal for the ADE series.

For computational purposes, it is often useful to have an expression for
the basis \ebasis\
which makes its relationship with the eigenvectors of the
Cartan matrix explicit. For each exponent $s$, define
\eqn\coxdefs{\eqalign{a_\B^{(s)}&=\sum_\B q_i^{(s)}\a_i\qquad
                      a_\W^{(s)}=\sum_\W q_i^{(s)}\a_i\cr
                      l_\B^{(s)}&=\sum_\B q_i^{(s)}\l_i\qquad
                      l_\W^{(s)}=\sum_\W q_i^{(s)}\l_i.\cr }}
Then, provided the eigenvectors of the Cartan matrix have unit length,
the vectors $a_\B ,a_\W$ are unit vectors while $|l_\B |=|l_\W
|=1/2\sin\t_s$, where $\t_s=s\pi /h$, and they enjoy a number of other
properties, including
\eqn\coxprops{\eqalign{a_\B^{(s)}\cdot a_\W^{(s)}&=-\cos\t_s,\quad
l_\B^{(s)}\cdot l_\W^{(s)}={\cos\t_s\over 4\sin^2\t_s}\cr
a_\B^{(s)}\cdot l_\W^{(s)}&=0=a_\B^{(s)}\cdot l_\W^{(s)},\quad
a_\B^{(s)}\cdot l_\B^{(s)}=a_\W^{(s)}\cdot l_\W^{(s)}={1\over 2}.\cr}}
The eigenvectors of the Coxeter element can then be
written in terms of these; for example, a convenient choice is
\eqn\coxeigen{e_s=\rho_s(a_\B^{(s)}+e^{is\pi
/h}a_\W^{(s)}),}
where $\rho_s=1/\sqrt{2}\sin\t_s$. The choice of
normalisation and the condition
\eqn\econj{e_{h-s}=e_s^*,}
require
\eqn\ashminuss{a_\B^{(h-s)}=a_\B^{(s)}\quad a_\W^{(h-s)}=-a_\W^{(s)}}
and
\eqn\qprop{\eqalign{q^{(s)}_\B&=q^{(h-s)}_\B\cr
           q^{(s)}_\W&=-q^{(h-s)}_\W .\cr}}
Armed with these facts, it is straightforward to obtain a representation
of roots or weights in this basis; for example, the fundamental weights
have components  given by
\eqn\weightcomps{\l_k^{(s)}=\cases{\rho_sq^{(s)}_k,&if\ $k \in \B$\cr
\rho_sq^{(s)}_ke^{is\pi /h},&if\ $k \in \W .$\cr }}

The masses of the affine Toda theory are proportional to the components of
$q^{(1)}$. Moreover, assuming
the other classically conserved quantities are preserved in the quantum
theory and are compatible
with the bootstrap \pbootstrap\ leads to the
conclusion \refs{\rKMa ,\rDc}
that the
single-particle states in the quantum theory are eigenstates of the
quantum operators $P_{s+kh}$,
with eigenvalues related  to the eigenvectors ${\bf q}^{(s)}$
of the Cartan matrix. Invariance under parity requires $p^a_r=p^a_{-r}$
and the fact that all fusing angles appearing in \pbootstrap\ are
multiples of $\pi /h$ requires $p^a_s=p^a_{s+2kh}$ where $k$ is any
integer. Together, these two requirements imply:
\eqn\prels{p^a_{s+2kh}\propto q_a^{(s)}\qquad p^a_{s+(2k+1)h}\propto
q_a^{(h-s)}.}
Hence, using \qprop\ above,
\eqn\prelsa{\eqalign{&p^\B_{s+kh}\propto q_\B^{(s)}\cr
                     &p^\W_{s+kh}\propto (-)^kq_\W^{(s)}.\cr}}
(Note, $s$ will always be taken to lie in the range $1,\dots ,h$.)

For spin $\pm 1$, the conserved charges are the light-cone momentum
components and the eigenvalues are just the masses.
The particles are distinguished from each other using the conserved
quantities but the single particle states will be
labelled by their momenta only, all other
labels being suppressed for convenience of notation.

For later use, a couple of alternative expressions for the
S-matrices for the various affine Toda theories will be given, each of them
equivalent (with the proviso noted below)
to the expressions provided in \rDc . Here,
as there and in earlier works \refs{\rBCDSb ,\rBCDSc}, the
block notation will be adopted in which the basic element of any of the
conjectured S-matrices is constructed from the
element
\eqn\block{(x,\T )_+ =\sinh\left({\T\over 2}+{i\pi x\over 2h}\right)}
where $\T$ is the rapidity difference for the process and $x$ is an
integer. The $\T$ dependence is made explicit here
for reasons which will become apparent in the next section.
  The basic building block itself is then defined to be
\eqn\sblock{\{ x, \T\}_+ = {(x-1, \T )_+(x+1, \T )_+\over (x-1+B, \T )_+(x+1-B,
\T )_+}}
where the function
$$B(\beta)={1\over 2\pi}\, {\beta^2 \over 1+\beta^2/4\pi}$$
contains the conjectured coupling constant dependence.
The first expression, given in \rDd , can be summarised as
follows:
\eqn\slambda{S_{ab}(\T )=\prod_{p=1}^h\{
2p+1,\T\}_+^{\lambda_a\cdot w^{-p}\phi_b}\qquad \T =\t_a -\t_b}
at least provided the two particles $a$ and $b$ share the same colour. If
the two particles correspond to different colours then the appropriate
expression is \slambda\ but the particle labelled $\W$ has its rapidity
effectively incremented by $i\pi /h$. In other words, the appropriate
S-matrix elements are obtained by replacing the rapidity $\T$ by
\eqn\thshift{\T_{\W\B}=\T +i\pi /h \quad\hbox{or}\quad \T_{\B\W}=\T -i\pi
/h.}
There are other, equivalent expressions in terms of the \lq unitary' block
$$\{ x, \T\}_+/\{ -x, \T\}_+,$$ but they are not so useful here.
The corresponding
minimal S-matrix is obtained from \slambda\ by simply deleting any
$\beta$ dependent term in \sblock .

These S-matrices are unitary, satisfy crossing requirements and fulfil the
bootstrap conditions on the odd order poles. They are analytic in the
rapidity difference $\T$ with $\beta$-independent poles on the physical
strip ($\hbox{Im}\T \ \in [0,\pi ])$. The poles may be
of quite a high order but appear to be compatible with perturbation theory
as far as has been checked \rBCDSe .

It will be convenient in the next section to consider a two (complex)
dimensional space of rapidity variables, $\t$ and $\bar\t$, corresponding
to a complexification of the light-cone momentum variables $p_{\pm}$. A
representation of the exchange relation will be constructed in this
complex space in the first instance and subsequently restricted to a
\lq physical'
submanifold. In part, this is motivated by an analogy with conformal field
theory where the two variables $z$ and $\bar z$, on which all conformal
fields depend, are often treated as independent complex coordinates with a
restriction to the euclidean section $z^*=\bar z$ left to a late stage of
a calculation. In the present case, it is the mass-shell condition
$p\bar p=p_+p_-=m^2,\hbox{or}\ \bar\t =-\t$ which selects the physical
submanifold. Note, the full vertex is automatically an analytic function
of $\t$ on this submanifold, in contrast to the conformal field theory
situation in which a typical (non-chiral) vertex operator is the product
of a function of $z$ and a function of $z^*$ on the euclidean section. A
feature of this kind is clearly needed---the S-matrix is itself
analytic in $\t$,
while typically, correlation functions in conformal field theory are not.

Consider the minimal S-matrix, in which the factors containing the
coupling constant dependence are omitted. There
are a number of continuations of \slambda\ off the $\bar\T =-\T$
sub-manifold. Here, just one will be given:
\eqn\srearrange{S^{\rm min}_{ab}(\T_{ab} ,\bar\T_{ab} )
={F_{ab}(\T_{ab},\bar\T_{ab})\over
                                     F_{ba}(\T_{ba},\bar\T_{ba}) }}
where,
\eqn\Fdef{F_{ab}(\T_{ab},\bar\T_{ab})={\prod_{p=1}^h
(-2p,\T_{ab})_+^{\l_a\cdot w^{-p}\l_b}\over \prod_{p=1}^h
(-2p,\bar\T_{ab})_+^{\l_a\cdot w^{-p-1}\l_b}},}
at least when the two particles share the same colour. When the colours
are different, {\bf both} $\t_\W$ {\bf and} $\bar\t_\W$ are shifted by
$-i\pi /h$. Note, for $\t_\W$
this is the opposite sign to the shift appearing in
the previous formula \thshift .
To check agreement between \srearrange\ and \slambda\ (when
$\bar\T_{ab}=-\T_{ab}$),
the following inner product identities are useful:
\eqn\wtprods{\eqalign{\l_\W\cdot w^{-p}\l_{\W}\p
=\l_{\W}\p\cdot w^{-p}\l_\W
&\qquad \l_\B\cdot w^{-p}\l_{\B}\p
=\l_{\B}\p\cdot w^{-p}\l_\B\cr
\l_\W\cdot w^{-p}\l_{\B}
&=\l_{\B}\cdot w^{-p-1}\l_\W .\cr
}}
It should be noted that whereas \slambda\ is a meromorphic function of
$\T_{ab}$ this is not generally true for the expressions occuring in
\Fdef ; since the inner
products of weights are not usually integers, these functions
will have individually a complicated cut structure.

\newsec{Representing the exchange relation}

The basic ingredients of the construction will be described first and then
elaborated to provide a representation of the exchange relation
for the minimal S-matrix.

For each fundamental weight $\l$ define a string-like, rapidity dependent
field as follows:
\eqn\lstring{X^\l (\t )=\sum_{r=s+kh}{h\over
r}e^{-r\t}\l^{(h-s)}c_{r}.}
In \lstring , the sum extends over all integers $k$ and exponents $s$ and
the Fock space annihilation and creation operators satisfy the commutation
relations
\eqn\ccomm{[c_{r},c_{r\p}]=(r/h)\delta_{r+r\p ,0}.}
The field $X^\l$ is periodic in $\t$ with period $2\pi i$ and a Coxeter
rotation of the weight is equivalent to a shift $2\pi i/h$ in $\t$. In
that sense, the field is \lq twisted'. There is a ground state satisfying
$$c_r|0>=0\qquad r>0.$$

The commutation relation between the annihilation part of a field
$X^\l_+(\t )$ and the creation part of another, similar, field
$X^{\l\p}_-(\t\p )$
is given by:
\eqn\xcomm{\left[ X^\l_+(\t ),\, X^{\l\p}_-(\t\p
)\right]=\sum_{p=1}^h(\l\cdot w^{-p}\l\p )\ln \left( 1-w^pe^{\t\p
-\t}\right),}
at least provided Re$\t >\hbox{Re}\t\p$.
Define a vertex operator to be the normal-ordered exponential of
such a field
\eqn\defvert{V^\l (\t )=\ts :\exp X^\l (\t ):\ts \equiv e^{X^\l_-(\t
)}e^{X^\l_+(\t )}.}
Then, the product of two
vertex operators can be normal-ordered producing an
extra factor depending upon the rapidity difference:
\eqn\reorder{V^\l(\t )V^{\l\p}(\t\p )=\prod_{p=1}^h\left( 1-\exp (\t\p -\t
+2\pi ip /h)\right)^{\l\cdot w^{-p}\l\p}:V^\l(\t )V^{\l\p}(\t\p ):,}
provided the rapidities bear the above relation to each other. The factor
on the right hand side of \reorder\ is strikingly similar to the factors
appearing in the S-matrix formulae \srearrange\ and \Fdef , but not the same.
A similar calculation with the ordering of the two operators reversed
produces the same factor (via manipulations valid in the complementary
region Re$\t <\hbox{Re}\t\p$ and up to a possible phase factor independent
of rapidity). Hence, a comparison of the two orderings after analytic
continuation in rapidity provides a trivial exchange relation.

The vertex operator introduced above is a conformal operator with respect
to the Virasoro generators built from the $c$-Fock space operators:
$$L_n={1\over 2}\sum_r c_rc_{n-r}.$$
This
observation explains the feature just described.
To obtain an exchange relation corresponding to the massive affine
Toda field theories the vertex operator will need to be \lq delocalised'
and its conformal nature destroyed.

A second remark concerns the commutation relation of a Fock space operator
with the vertex:
\eqn\cvcomm{\left[c_{s+kh},\, V^\l (\t )\right]=\l^{(s)}\, e^{(s+kh)\t}\,
V^\l (\t ).}
Recalling the earlier observations \weightcomps\ and \prelsa , it is tempting
to regard the
annihilation operators in the Fock space as the conserved charges of the
affine Toda theory, and take the vertex operators (suitably modified) to
describe
the single particle states. Indeed, the  $-i\pi /h$ shift in
rapidity for a type $\W$ particle, bearing in mind \prelsa\ and \weightcomps ,
renders \cvcomm\ compatible with this assumption for either of the two
colours. This rather natural point of view is the one adopted here, at
least
tentatively, bearing in mind that the relationship between this
representation of the conserved quantities and the usual one in terms of
the fundamental fields of the theory is missing, and that there is
certainly a subtlety to be understood with regard to the hermiticity of
the operators concerned; the conserved quantitities are given classically
as real functionals of the fields and their derivatives.

Since, as remarked earlier, there are two
mutually commuting sets of
conserved quantities of opposite spin there will have to be at least one
extra set of Fock space operators. These are quite naturally associated
with the variable $\bar\t$ which will, as mentioned above,
be regarded as independent of $\t$
for most computational purposes. The conserved quantities with the
opposite spin nevertheless share the same eigenvalues on the particle
states apart from their rapidity dependence. Thus, if an extra set of
Fock space operators $\bar c_r$ is introduced they and their associated
vertex operator $\bar V^\l (\bar\t )$ will be expected to satisfy
\eqn\cbarvcomm{\left[\bar c_{s+kh},\, \bar V^\l (\bar \t )\right]
=\l^{(s)}\, e^{(s+kh)\bar\t}\,
\bar V^\l (\bar\t ),}
at least for $r>0$. Note, this expression is also compatible with the
shift in $\bar\t$ for a type $\W$ particle.

Introducing a second set of operators also provides
the opportunity for some \lq delocalisation' in the following sense. The
conformal nature of the vertex operator can be destroyed by shifting the
rapidity dependence in the annihilation part of the vertex (ie without
upsetting \cbarvcomm ) relative to that in the creation part. Indeed, a
shift of $2\pi i/h$ seems to be exactly what is required to produce the
minimal S-matrix \slambda .
Thus, introducing the field $\bar X^\l (\bar\t )$, it is convenient to set
\eqn\defshvert{\bar V^\l (\bar\t )=e^{\bar X_-^\l (\bar\t )}e^{\bar
X_+^{w\l}(\bar\t )}}
and straightforward to repeat the normal-ordering calculation, giving
\eqn\reorderbar{\bar V^\l(\bar\t )\bar V^{\l\p}(\bar\t\p )=\prod_{p=1}^h
\left(1-\exp (\bar\t\p -\bar\t
+2\pi ip /h)\right)^{\l\cdot w^{-p-1}\l\p}:\bar V^\l(\bar\t )
\bar V^{\l\p}(\bar\t\p ):,}
this time valid in the region $Re\bar\t >Re\bar\t\p$. Unfortunately, this
is not quite what is required for \Fdef , the exponent in \reorderbar\ has
the wrong sign. To put that right, the signature of the barred Fock-space
operators should be reversed:
\eqn\cbarcomm{[\bar c_r,\bar c_{r^\prime}]=-(r/h)\delta_{r+r^\prime ,0}.}

The two calculations \reorder\ and \reorderbar\ indicate that a vertex
operator associated with a particle of type $\B$ could be
\eqn\bvertex{V^a(\t ,\bar\t )=V^{\l_a}(\t )\bar V^{\l_a}(\bar\t )}
while that associated with a particle of type $\W$ is a similar expression
but with $\t$ and $\bar\t$ each shifted by $-i\pi /h$.
The latter is required to match the slight
difference in the
formulae for the S-matrices for the two types, mentioned earlier. Adopting
the composite vertex operator and putting together the reordering effects
gives
\eqn\norder{V^a(\t_a,\bar\t_a) V^b(\t_b,\bar\t_b)
=F_{ab}(\T_{ab},\bar\T_{ab}): V^a(\t_a,\bar\t_a) V^b(\t_b,\bar\t_b):}
where $F_{ab}$ is defined in \Fdef .
Repeating the calculation in the opposite order gives a similar expression
to \norder , but with labels $a,b$ reversed. Assuming an analytic
continuation into a common region of complex rapidity and comparing the
two reordering expressions yields the exchange relation, \braid\ for the
minimal part of the S-matrix.

Having achieved a representation of the minimal S-matrix, the same ideas
may be adapted to introduce the coupling constant dependence. One way to
do so is to introduce
another pair of string-like fields $Y^\l (\t)$ and $\bar Y^\l (\bar\t )$,
with corresponding sets
of annihilation and creation operators $d_r,\,\bar d_r$,
together with corresponding vertex operators $W^\l (\t )$ and $\bar W^\l
(\bar\t )$ defined by
\eqn\ydefs{W^\l (\t )=:e^{Y_-^\l (\t )}e^{Y^\l_+(\t -i\pi (2-B)/h)}:\qquad
          \bar W^\l (\bar\t )=:e^{\bar Y_-^\l (\bar\t )}
e^{\bar Y^{w\l}_+(\bar\t -i\pi (2+B)/h)}:.}
Note, both constituent vertex operators are \lq delocalised' this time
by an amount dependent upon $B(\beta )$. Note also, the commutation
relations of the $d,\bar d$ Fock space operators
have the same signature as the $c,\bar c$ operators, respectively.

The operator representing a particle of type $\B$
is now taken to be
\eqn\fullvertex{V^a (\t_a ,\bar\t_a )=V^{\l_a} (\t_a )\bar V^{\l_a}
(\bar\t_a )
W^{\l_a} (\t_a )\bar W^{\l_a} (\bar\t_a ).}
Effectively, the four sets of annihilation and creation operators can be
combined into a four-dimensional vector in a space with a metric of signature
$(++--)$. This fact is reminiscent of a comment by Ward \RF\rWc\Wc\
concerning the
embedding of an affine Toda field theory in a self-dual gauge theory.
This construction is certainly ad hoc and there may be other, more
subtle, ways of achieving the same goal. Nevertheless, the example given
establishes that the exchange relation can be represented in
principle.

Returning to the vertex operators it is worth emphasising that the
coefficients of the Fock space creation operators in \lstring\ are
proportional to the eigenvalues of the conserved quantities. This follows,
irrespective of the colour of the particles, from the relations
\weightcomps\ and \prels , taking into account the relative rapidity shift
$-i\pi /h$ for the type $\W$ particles.
In other words, the coefficients in \lstring\ may be
taken to be the conserved charges, up to a factor that may depend upon
$\beta$ and the spin, but not on the particle type. Thus for any particle
$a$, regardless of type, the string-like field may be written
\eqn\newstring{X^a(\t )=\sum_{r=s+kh}{h\over r}e^{-r\t}\pi^a_{-r}c_{r},}
where
$$\pi^a_{s+kh}\propto p^a_{s+kh}.$$
This remark is very
interesting in the context of the fusing relation \opprod , since it
implies that the fusing relation for the operators
follows from the bootstrap for the eigenvalues of the conserved charges.
To see this, consider a normal-ordered product of vertex operators for a
pair of particles $a$ and $b$,
such as that appearing on the right hand side of \reorder , and evaluate
it for the values of rapidity appropriate to a fusing process \opprod .
Thus, it is required to evaluate
\eqn\vfuse{\eqalign{
:V^a(\t_{c}-&i\bar U^b_{ac})V^b(\t_{c}+i\bar U^a_{bc}):\cr
&=:\exp\[ \sum_{r=s+kh}{h\, e^{-r\t_{c}}\over
r}\(\pi^a_{-r}e^{ir\bar
U^b_{ac}}+\pi^b_{-r}e^{-ir\bar U^a_{bc}}\)c_{r}\]:.\cr}}
However,
using the conserved quantity bootstrap \pbootstrap , valid for any spin $r$,
and remembering the $\pi$'s and $p$'s are proportional with a factor
independent of
the species of particle, the right hand side of equation \vfuse\ is precisely
$V^{c}(\t_{c}),$
as desired. The same calculation works for all the other pieces of any
vertex operator, since the \lq delocalisation' is independent of the
particle type, and hence for the complete operators $V^a(\t_a ,\bar\t_a)$
defined in \bvertex\ or \fullvertex .

It remains to study the singularity structure of the normal-ordered
operator product \norder , at least when the particles are on-shell so that
$\bar\t =-\t$. Only brief comments will be made here, restricted to the
minimal case, a fuller analysis will be given elsewhere. Firstly, although
the numerator and denominator on the right hand side of \Fdef\ have branch
cuts (since the inner products of weights are not usually integers), when
$\bar\t =-\t$ these combine nicely to yield poles and zeroes. Moreover,
the poles and zeroes also appear in the S-matrix, unless they cancel in
the exchange relation. There are just two types of cancelling poles and
zeroes; zeroes at $\T_{ab} =0$ and poles at $\T_{ab} =i\pi$,
the latter occuring
only when $b=\bar a$, the former for $b=a$\rlap.\footnote{**}{Note, the
vertex operator $V^{*a}(\t_a, \bar\t_a)$, obtained from \fullvertex\ by
reversing the sign of the weight $\lambda_a$ on the right hand side, is
in a sense, the operator conjugate to $V^a (\t_a ,\bar\t_a )$; it
satisfies the same exchange relation, it has a commutator with the
conserved quantities with the opposite sign to \cvcomm ,
and there is a simple pole
at $\t_a =\t_a^\prime$ in the operator product of $V^a (\t_a ,\bar\t_a
)V^{*a} (\t^\prime_a ,\bar\t^\prime_a )$} The coefficient of the pole at
$\T_{a\bar a} =i\pi$ in the $a\bar a$ operator product is precisely unity.
Otherwise, the poles come with associated zeroes; in other words,
$F_{ab}$ is a product of factors
of the form
$${(x,\T )_+^n\over (-x, \T )_+^m},$$
where $m=n \ {\rm or}\ n\pm 1$. If $m=n$, then the poles and zeroes are
destined to be even order poles and zeroes in the exchange relation and
ought not to participate in the bootstrap; if $m=n+1$, the pole indicates
the presence of a forward-channel bound state and participates in the
bootstrap; if $m=n-1$ it does not. In other words, it is the algebraic sum
of poles and zeroes that appears to be relevant for the fusing relation,
not the mere existence of poles. Except by examining the S-matrix, it is
not clear why this should be so. For a deeper analysis of the S-matrix
pole structure, including an explanation of some of the above remarks,
see ref\rDd .

\newsec{Discussion}

Despite the lack of any construction in terms of the elementary Toda
fields, the vertex operators presented here do at least provide a succinct
summary of the affine Toda S-matrices and elucidate further the algebraic
structure of these theories. Moreover, given \opprod , the S-matrix
expressions \slambda\ automatically satisfy \bootstrap , a fact which
otherwise needed to be checked separately. A particularly nice feature of
the full vertex operator, expressed in terms of fields \newstring , is its
dependence (without distinction of colour), on the
eigenvalues of the conserved charges. These statements apply equally well to
the full S-matrix or its minimal part, the vertex representation of the
former being admittedly the more ad hoc. An important ingredient in the
construction presented here is the delocalisation of the vertex operators
as a means to obtain the necessary breaking of conformal symmetry. This is
fairly arbitrary,
though in fact reminiscent of a vertex operator construction of
quantum affine algebras, due to Frenkel and Jing \RF\rFJa\FJa .

At a technical level, the problem of performing the analytic continuations
to make the comparison between the two halves of the operator reordering
calculations in \norder\
is quite delicate because of the branch points in the terms  in \Fdef ;
there may be  rapidity-independent phases to be
taken care of by the introduction of matrix factors in the vertex
operators (see, for example, \rLa ). In fact, the construction presented
here is \lq heterotic', being
asymmetrical between the terms depending upon $\T$ and $\bar\T$.
Indeed, as remarked earlier, the part of the exchange relation arising
from \reorder\ is merely a phase independent of $\T$. However, it plays
an important r\^ole at intermediate stages, ensuring that the poles in
the fusing relation \opprod\ have integer powers. This and the desire to
have quantities to represent all (positive and negative) spin conserved
charges motivated the inclusion of an apparently redundant set of
Fock space operators. It is also possible they serve to eliminate
the need for \lq cocycle' factors of the type mentioned above.

Finally,
it is worth remarking that the construction of classical soliton solutions
in terms of $\tau$-functions \RF\rMJa\MJa\ involved vertex operators of a
similar type, although in the context of affine
Toda theory a complete
discussion of solitons  is not yet available \RF\rOTb{\OTb\semi\OTc\semi\Hc}.
It will be interesting to see how the classical information
is related to the real $\beta$ scattering theory.
\vfill\eject
\bigskip\centerline{\bf Acknowledgements}
\smallskip
We are grateful to D. Bernard for bringing ref\rFJa\ to our attention,
to F. Smirnov and R. Sasaki for useful comments and to the Research
Institute for Mathematical Sciences, Kyoto University for a fruitful
visit.
One of us (E.C.) wishes to thank the Institute of Theoretical Physics at
the
University of Santa Barbara, and the organisers of the Workshop on
Conformal Field Theory for their hospitality during the initial stages of
this work. The other (P.E.D.) is grateful to the Royal Society for a
Fellowship under the European Science Exchange Programme.
The research was
supported in part by the National Science Foundation under grant No.
PHY89-04035, supplemented by funds from the National Aeronautics and Space
Administration, at the University of California at Santa Barbara.

\listrefs
\end